# The Alan Turing Institute

December 2023

# Exploring responsible applications of Synthetic Data to advance Online Safety Research and Development

Pica Johansson, Dr Jonathan Bright, Dr Shyam Krishna, Claudia Fischer, Prof David Leslie

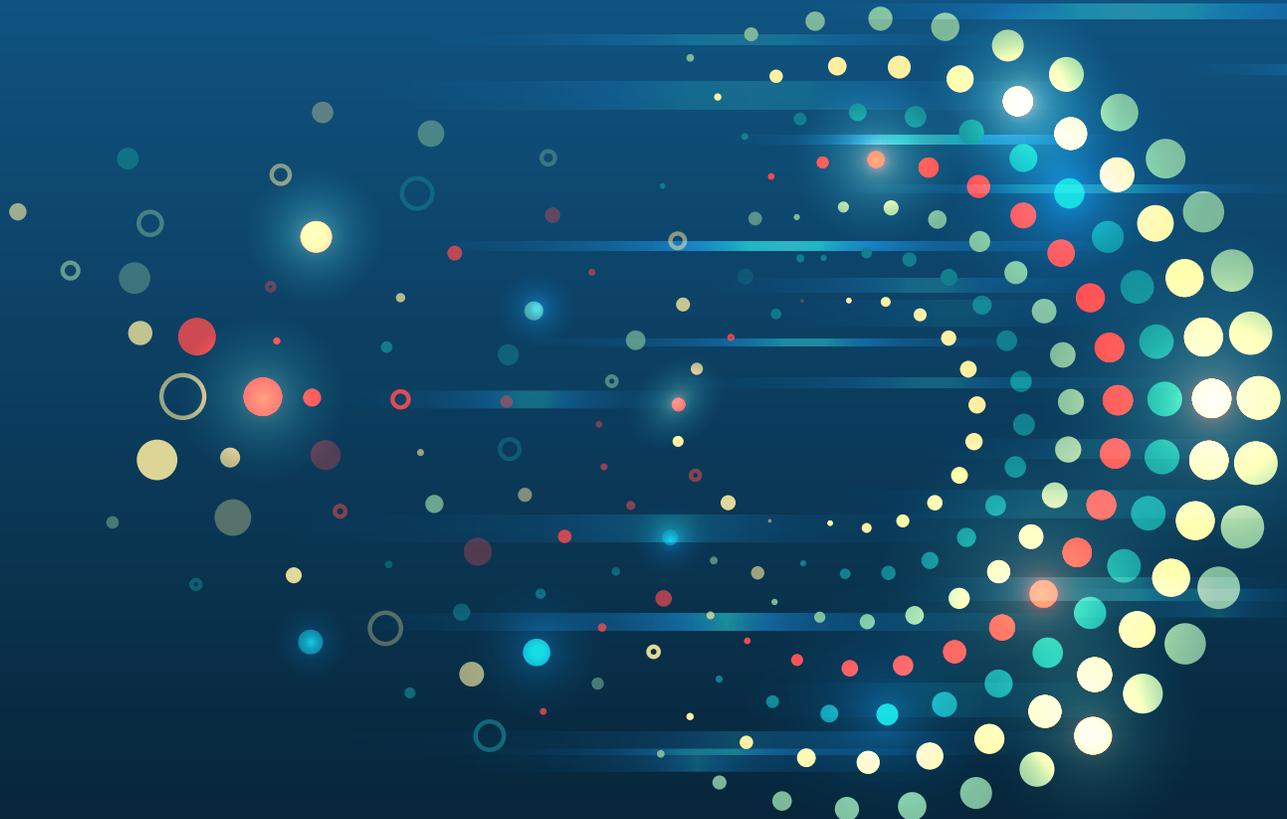

# Table of Contents

## Authors

**Pica Johansson**

Research Associate in the Online Safety team at The Alan Turing Institute.
https://www.turing.ac.uk/people/researchers/pica-johansson

**Dr Jonathan Bright**

Head of Online Safety and Head of AI for Public Services at The Alan Turing Institute.
https://www.turing.ac.uk/node/1794

**Dr Shyam Krishna**

Postdoctoral Research Associate in the Ethics and Responsible Innovation team at the Alan Turing Institute.
https://www.turing.ac.uk/people/researchers/shyam-krishna

**Claudia Fischer**

Research Associate in the Ethics and Responsible Innovation team at the Alan Turing Institute.
https://www.turing.ac.uk/people/researchers/claudia-fischer

**Prof David Leslie**

Director of Ethics and Responsible Innovation Research at The Alan Turing Institute and Professor of Ethics, Technology and Society at Queen Mary University of London.
https://www.turing.ac.uk/people/researchers/david-leslie

## Acknowledgements

We thank Christian Butterworth, Thomas Stukings, Harry Riley and Jo Brooks from the UK Department for Science, Innovation and Technology for their collaboration on this report. We are grateful to Dr Yi-Ling Chung and Hannah Rose Kirk at the Alan Turing Institute as well as members from Ofcom's Technology Policy team for their support in the ideation stage of this report. We would also like to thank Mark Durkee and Dave Buckley from the Centre for Data Ethics & Innovation for their thoughtful feedback on the final report. From the Alan Turing Institute we would also like to thank John Francis and Dr Florence Enock for their valuable feedback on the draft report and Joudy Bourghli for her help designing the report layout.

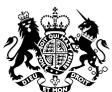
Funded by UK Government

# Recommendations for Practitioners Using Synthetic Data to Accelerate Online Safety Research and Development

The use of synthetic data provides an opportunity to accelerate online safety research and development efforts while showing potential for bias mitigation, facilitating data storage and sharing, preserving privacy and reducing exposure to harmful content. However, the responsible use of synthetic data requires caution regarding anticipated risks and challenges. With this in mind, we have created a set of recommendations to provide guidance on how practitioners and innovators can use synthetic data in a responsible way. These recommendations are based on the information provided in the preceding sections of the report and are subject to change as new research and best practices emerge.

- **The use of synthetic data presents a unique opportunity for teams to expedite their research and development efforts, while simultaneously mitigating biases, facilitating data storage and sharing, preserving privacy, and reducing individuals' exposure to harmful content.** However, responsible innovation requires careful consideration of the potential risks and challenges associated with this technology. In order to approach innovation with responsibility, teams can leverage the many benefits of synthetic data to create more effective and safer R&D tools.

- **Teams should justify the use of synthetic data and make plans for storage and evaluation before generation.** This involves documenting a careful assessment of the benefits and drawbacks of using synthetic data for a particular project or use-case, as well as justifying why existing data is insufficient. By defining the intended use-case and desired outcomes at the outset, teams can prevent mission creep and reduce the risk of using the data in projects it was not originally intended for. Although synthetic data offers many efficiencies, its responsible use requires specific resources and skill sets, such as for storage and evaluation, which must be planned for accordingly before starting the project.

- **When generating and using synthetic data it is crucial to track its provenance.** To ensure proper provenance tracking, teams are recommended to document where the synthetic data comes from and how it was used to create further data points. This safeguards against creating feedforward supply chains where synthetic data is used to generate more synthetic data. Such chains can dilute, distort, or compromise the underlying data distributions, leading to unreliable results. This information should be recorded in clear documentation that can be easily accessed and reviewed. By establishing and adhering to provenance tracking guidelines, researchers can maintain the quality and integrity of their synthetic data, and ensure that their results are trustworthy and reliable.

- **When deciding how to measure data leakage and privacy-preserving properties of a dataset, it is necessary to consider the type of data and its uses.** Although synthetic datasets hold the promise of being more privacy-preserving than their non-synthetic counterparts, they are not privacy-preserving by default, and careful analysis is required when measuring the levels of data leakage of a particular dataset. Creating multiple synthetic datasets to answer different questions instead of only one or using differential privacy in conjunction with synthetic datasets are some avenues to enhance the privacy of the data. This is especially important when dealing with non-tabular data whose privacy-preserving properties are even more difficult to measure, in large part because the concept of 'similarity' between data points is not straightforward in these cases (i.e., two text data points can be different yet convey the same meaning).

- **As such, there is no consensus on the one way of measuring data leakage.** Therefore, when trying to determine how private a dataset is, it is important to consider the type of data being generated, the different methods used to measure privacy levels, and what the best practices to enhance a dataset's privacy in that particular context are.

- **Synthetic data does not present a silver-bullet for de-biasing datasets; models used to generate synthetic data, as well as their output, must be assessed for biases.** Although synthetic data can help mitigate bias in non-synthetic datasets by generating synthetic data points to increase the representation of underrepresented groups, there's still a risk that the synthetic data itself may be biassed. For instance, it might preserve aggregate characteristics, or the models used to generate the data may be biassed themselves. While some research suggests ways to detect and mitigate bias, others argue that such measures are insufficient. Therefore, it's important to document bias-correction measures and assess how they might be impacting the dataset. By doing so, teams can ensure that they are using synthetic data in a responsible and effective manner.

- **The release of synthetic data must be controlled.** While synthetic data shares many of the same risks as the original dataset has in terms of public or semi-public release or publication, datasets used for online safety purposes might pose even higher risks if they contain a large proportion of potentially harmful content. In such cases, the data could be used for malicious purposes, such as generating and spreading further harmful content, if it falls into the wrong hands. Therefore, it is essential to establish clear guidelines for the release of synthetic data and ensure it is only shared with trusted parties who have legitimate need for it.

- Finally, the use of synthetic data as training data for models should be approached cautiously. While there is considerable potential for synthetic data to enhance the scale and variety of examples in a training dataset, there is also increasing concern that model quality progressively declines when using exclusively (or even partially) synthetic datasets.

# Introduction

Data are the foundational components used to train, test, and validate machine learning applications. As demand for machine learning technologies spreads across more and more areas of human activity and data sources multiply as a result of extensive digitisation, the challenges surrounding the responsible sourcing of large, high quality datasets have grown more acute and widespread. These challenges are even more pronounced in the case of the collection, management, and use of online safety data where issues around the privacy and sensitivity of digital trace data, and potential vicarious harms to both annotators and data subjects, are extremely complex.

The difficulties surrounding the collection of data for machine learning have led to a recent wave of interest in how to move beyond traditional data gathering and preprocessing approaches towards other techniques that can leverage more impact from existing data, and reduce the burden of data collection and preparation. One of these techniques is called *synthetic data generation*: a process by which novel, fabricated data are created, either by humans or by machines, to imitate a genuine dataset.

To take a simple example, imagine needing a large dataset of photos of sunsets, each varying slightly in their colours, setting and location. One approach would be to source these photos from genuine photographers, however this might prove time consuming and expensive. A second approach would be to collect a small amount of photos, and then use these to train a generative model for creating synthetic data. This model would allow you to use your existing photos as input to create a much larger set of photos that capture the general composition of the ones you already have by replicating them with random variations learned from the existing photos, and not just by duplicating them.

Synthetic data generation presents a number of opportunities. It allows augmentation of existing datasets; an increase in both the quantity and diversity of data examples; and novel possibilities for mitigating biases and improving the balance and representativeness of original training datasets (for example, by adding data that represent members of minority groups who have not been properly included in the data collection process). It can allow for quick prototyping and testing of systems. Synthetic datasets also allow the potential fabrication of as yet unseen or hypothetical scenarios, which could help improve system robustness. In theory, synthetic data should also be able to facilitate more responsible data sharing insofar as risks of compromising privacy and potential data leakage are mitigated by the generation of non-identifiable synthetic datasets.

> Synthetic data has generated much optimism and has already started to be applied widely across different domains in machine learning, with Gartner suggesting that already 60% of data used to train models will be synthetic by 2024 (White, 2021).

However, as an emerging field, the range of ethical risks and challenges associated with using synthetic data is yet to be fully explored. The aim of this report is to critically examine the ethics of using synthetic data, with a particular focus on its use in online safety technology, for example the technologies used in the detection of harmful content online. The report first outlines the context within which ethics will be discussed by defining key concepts, application areas and providing a brief overview of how synthetic data are created. Next, this report details the ethical implications associated with creating synthetic data, sharing synthetic data and modelling with synthetic data. Throughout the report, we propose holistic and practical approaches to governing synthetic data, aimed at policymakers and other stakeholders.

The report first outlines the context within which ethics will be discussed by defining key concepts, application areas and providing a brief overview of how synthetic data are created. Next, this report details the ethical implications associated with creating synthetic data, sharing synthetic data and modelling with synthetic data. Throughout the report, we propose holistic and practical approaches to governing synthetic data, aimed at policymakers and other stakeholders.

# Background

Synthetic data are data that have been created to represent or mimic the properties of underlying 'real' or 'genuine' data. The idea of synthetic data is not a novel one (Nikolenko, 2021; Raghunathan, 2021), and indeed in the area of tabular data (such as data created by surveys) its study is well advanced (Jälkö et al., 2021). However, recent advances in artificial intelligence (AI) have meant that synthetic text, audio, video and image data have also become easier to create, leading to an increasing amount of interest in the area and a wider variety of synthetic data types.

In this report, we place a focus on synthetic data that has been generated using a purpose built mathematical model or algorithm, with the aim of solving one or more data science tasks in the area of online safety (Jordon et al., 2022). We do not consider datasets that have been created entirely 'by hand'[1]. We also include in our discussion the generative models used to create these datasets.

Synthetic datasets aim for three properties: flexibility, fidelity, and privacy (Lin et al., 2019). A model allowing for high flexibility allows for greater variety in the data it produces, potentially scaling categories of data that are sparse in the original dataset. For example, flexibility in hate speech detection would mean that a dataset with a small proportion of various types of hate speech could be scaled to include many more different examples to train a more robust detection model. Fidelity refers to how closely the synthetic data replicates the underlying 'genuine' dataset. Privacy, finally, refers to how well the novel dataset that has been generated obscures the characteristics of the genuine data that underlie it. Fidelity and privacy may trade off with one another, since new synthetic data which is very similar to the 'genuine' data risks being less privacy preserving. However, a dataset of lower fidelity might come at the cost of its utility (though even very low fidelity data can have certain applications such as prototyping systems). The balance to strike between these different properties will be determined by the needs of the synthetic data's unique use-case.

Synthetic data can be created using many different types of models, in turn producing different types of data. This report does not intend to delve into the technical specifications of how to create synthetic data, however, we provide this brief overview for further reading:

- Synthetic image generation is often based on deep learning architectures, for example using Generative Adversarial Networks (GANs) (Goodfellow et al., 2020), Variational Auto-encoders (Kingma & Welling, 2022) and Diffusion Models (Ho et al., 2009). Diffusion models in particular power well known public image generation technology such as DALL·E [2] and Stable Diffusion[3].

---

[1] However, the techniques that we focus on in this paper all require some human input that influences their results as well as having a so-called 'human-in-the-loop'.

- A variety of techniques exist for the generation of synthetic tabular data, broadly based on first modelling a likely probability distribution on the basis of observed data, and then resampling from this probability distribution to create novel records (Nowok et al., 2016). More recently GANs have also been applied to its generation (Isasa et al., 2023; Marin, 2022).

- Synthetic simulations, for example of how users might react to a piece of harmful content in a news feed (Corbitt-Hall et al., 2019), may use agent-based modelling. Agent-based modelling is a family of techniques that involve modelling dynamics in systems through the creation of decision-making entities (called 'agents') and observing their behaviour on the basis of the interaction of simulated contexts, events and decision making rules (Bonabeau, 2002).

- Synthetic text generation is now often based on large language models (Adelani et al., 2020). Large language models power many of the models used in content moderation (Kerner, 2022), as well as well known recently released chatbots such as ChatGPT[4], BingChat[5] and Google's Bard[6].

The generation of synthetic non-tabular data such as text, audio, image and video is an area of considerable research interest. Compared to synthetic data that can be tabulated, these emerging technologies are more complex. However, as these types of data are used more widely in online content, interest in the production of models and products in this area has recently expanded, making it an important issue for online safety research and development to address. For instance, since 2022, industry-developed models that produce generative synthetic content have achieved widespread public attention. These include online services such as ChatGPT, a text-based chatbot, or text-to-image creators like Stable Diffusion, Midjourney[7], and DALL·E 2. It should be noted that these models, and the products built on them, are not deployed purely with the purpose of creating synthetic datasets. However, they showcase the power of generative AI technologies on which synthetic data relies, and the increasing ease with which novice users in the general public are able to engage with them to create texts, images and videos.

Synthetic datasets of varying levels of sophistication are regularly being published. While the performance metrics across types of data differ, generative technology has reached a point where text and images can be produced with an advanced level of fidelity. Producing audio and video has proven more difficult due to the complexity of factors involved in their generation. Models generating video require capturing of the spatial setting of a scene and visual information progressing across time, in addition to the more static visual presentation of the objects (Unterthiner et al., 2019). Similarly synthetic audio generation is challenging because of the complex need to regulate pitch, loudness and the effect of room acoustics (Engel et al., 2020). However, it is reasonable to expect continued technical advances in these areas.

---

[2] **https://openai.com/dall-e-2**
[3] **https://stablediffusionweb.com/**
[4] **https://openai.com/blog/chatgpt**
[5] **https://www.bing.com/new**
[6] **https://bard.google.com/**
[7] **https://docs.midjourney.com/**

> Despite the complexities, it is worth emphasising that synthetic data may be useful for the purposes of model training and evaluation even if it does not reach a level of realism which makes it indistinguishable from reality (i.e. it can have less than 100% fidelity).

A low-fidelity synthetic dataset that does not mimic all the attributes of the real dataset is still useful for understanding the characteristics of the dataset as a whole, and for establishing the context in which the dataset is being used. It can also enable system prototyping. This can help developers or policymakers understand the nature of phenomena they study without compromising the privacy of any individual data records in the dataset (Calcraft et al., 2021).

# Key application areas

The promise of using synthetic data has led to its application in many areas, but the derived benefits have largely been uniform across domains. Firstly, using synthetic data should enable easier data sharing and storage, as its handling is not subject to the same privacy requirements as authentic data. Synthetically generated data is also scalable and possible to augment to compensate for data scarcity or missing data points. The potential that synthetic data shows for being shared, stored, reused, and published in a privacy-preserving way has made it a popular alternative to real data in light of the growing demands for compliance with data privacy and protection legislation (Jordon et al., 2022). It should be noted, however, that synthetic data does not preserve privacy by default, as we will discuss further below.

The utility of synthetic data is also evident in its potential for augmentation and scaling. In practice, this means that once a model intended to produce synthetic data is built, it can provide large volumes of novel training data and generate diverse datasets with previously unseen examples, two factors critical for building new, resilient AI-powered models. However, it should be emphasised that synthetically produced data augmentation and scaling does not address real-world shifts in underlying data distributions that may occur after the collection of original training datasets, and hence cannot capture shifts or drifts in ground truth that come about through changes in measured populations over time.

Using synthetic data also allows for lower barriers to entry to innovate and to test prototypes and new solutions; and to do so responsibly. High fidelity synthetic datasets allow data scientists and engineers to gain a better understanding of existing systems, can help motivate better design choices and allow developers to test if new or proposed systems also work within a realistic simulation - for example in a digital sandbox environment[8] (Lin et al., 2019).

> Using synthetic data can therefore help stimulate technological innovation by enabling researchers, developers and other innovators to overcome entry barriers associated with data scarcity and to meet privacy and protection requirements.

Data scientists and engineers have found synthetic data particularly useful in a range of applications for the benefit of online safety, and the UK Government's Online Safety Data Initiative highlighted synthetic data as a potential solution to the data scarcity problem developers face when innovating for online safety tech (Online Safety Data Initiative, 2021). So far, synthetic data have helpfully been used to:

- Explore data in the development phase of new projects, as it enables researchers to train and finetune models ahead of accessing the real data. For example, synthetic data has been used to train models which create data used to better detect fake online reviews (Aghakhani et al., 2018; Dolhansky et al., 2020). Aghakhani et al. 's (2018) results showed that semi-supervised models using synthetic data produced results equivalent to state-of-the-art supervised learning.

- Explore models ahead of deployment, allowing for software testing and provisional models to be tested in a controlled environment. Large language models (LLMs), such as GPT4, Bard, LLaMA and Chinchilla are tested ahead of their release to mitigate against potential harmful outputs, such as offensive language and non-violent but unethical outputs. Results have shown that LLM-based red teaming is effective, as LLMs are seen to produce both diverse pieces of content (i.e. with high variance), and what is typically seen as difficult or 'edge case' content (Perez et al., 2022).

- Create synthetically generated sensitive data for abuse-detection, where real-data cannot be used or revealed (Wullach et al., 2021b, 2021a). These studies exemplified how synthetic data augmentation led to improvements across multiple hate speech datasets, preliminary driven by improvements in recall (sensitivity) as the researchers were able to generate one million examples – hundreds of thousands entries more than were previously available.

- Create data to train abusive language detection models in new types of mediums/modalities. For example, using synthetic audio to train models that can help detect abuse in transcripts of video games - an area that is difficult to study (ADL, 2021).

- Create data to help boost performance on classification tasks, such as hate speech detection. Researchers have for example effectively generated 274,000 synthetic data entries to cover implicitly toxic text at larger scale than is possible to generate by hand, providing an effective way to cover a wider range of demographic groups (Hartvigsen et al., 2022).

- Evaluate and benchmark models. For example, Zhou et al., (2021) created a synthetic dataset using a few-shot dialect translation system built with GPT3 (Brown et al., 2020) allowing them to "translate" tweets into an alternative phrasing with African American English dialectal markers. Using this synthetic data, the researchers tested how well a toxicity detection model performed on tweets with an alternative dialect, and automatised the required corrections to debias their toxicity-detection model.

- Create data that are sparse and difficult to collect. For example, create new data to challenge hate on platforms and/or respond with an alternative narrative, also known as 'counter speech' (Bonaldi et al., 2022; Chung et al., 2020; Fanton et al., 2021; Tekiroğlu et al., 2020; Zhu & Bhat, 2021). In Bonaldi et al. 's (2022) study, for example, the researchers were able to create examples of counter speech that were comparable to those created by humans, including at the same level of novelty, but at a fraction of the time humans would have required for the same task.

- Test and boost model performance for low-resourced languages. For instance, with the help of GePpeTto (an Italian version of GPT2), Chung et al. (2020) tested different resource conditions, including zero and few shot learning by using (i) synthetic machine translated data, (ii) real data, and (iii) their combination for generating counter speech. Simple augmentation, paired with language models, proved efficient for generating counter speech in Italian.

---

[8] A 'digital sandbox' can be understood as a virtual testbed for a selected number of projects where the project can be tested before being released into the 'real world' (European Institute of Public Administration, 2021).

These applications showcase how synthetic data can be used for online safety purposes, primarily adding efficiencies in speed, gained by the ability to generate many entries (often reducing human labour) and data quality, gained by increasing novelty and diversity in training data (contributing to better models)[9]. The improved detection of potentially harmful online content is facilitated by improved data quality, as the addition of synthetic data points pertaining to marginalised groups or under-represented demographics (that might be scarce in a genuine dataset) improves model performance.

> It is worth noting that despite the promise of large and varied quantities of data, the consequences of increasingly training machine learning models on synthetic data is yet to be fully understood. Initial research on this topic has found that once LLMs contribute much of the language to their own training they experience 'model collapse' (Alemohammad et al., 2023; Shumailov et al., 2023).

This is described as a degenerative process whereby such models forget the true underlying data distribution over time, even though there might not have been a shift in the distribution within the same time period. The value of data collected from genuine human interactions might therefore increasingly be valued, as we learn about the effects of developing, training and retraining models on synthetic data.

---

[9] Note however, that few application areas are seen to reap the benefits of the privacy preserving properties, as data related to online safety have sensitivities that go beyond data privacy and for that reason data sharing is highly restricted.

# Key ethical issues

The main focus of this report is to outline the ethical issues involved in use of synthetic datasets, particularly in the area of online safety. The following sections in the report discuss the key ethical issues involved in the use of synthetic datasets in online safety research and development. We structure our review around three areas: 1. The creation of synthetic datasets, 2. The sharing of synthetic datasets, 3. The use of synthetic datasets.

## Creating synthetic datasets

### Limiting potential exposure to harmful content

Within the AI project lifecycle, the way data classification and labelling is carried out has already raised ethical concerns regarding who views, handles, and classifies the data and how they are compensated (Hagendorff, 2022; Perrigio, 2023). Here, ethical considerations regarding how humans are involved in the synthetic data creation process must not be overlooked. This involvement can occur in two ways: using an initial real-world dataset with existing human-made labelling and classification for training, or involving humans in the crowdsourced labelling of an initially unlabelled real-world dataset or synthetic data to generate a training dataset (Jordon et al., 2022; Yiallourou et al., 2017). In both cases, consideration must be given to the individuals that label the data, how they are compensated and protected from vicarious harms from exposure to such data, who controls and manages these processes, and how downstream generated synthetic data is utilised.

Close and well-defined control of the training and experimental setup of synthetic data production pipelines is necessary to safeguard these individuals, and careful consideration must be given to how the synthetic data is used in wider real-world applications. For instance, in the context of managing potentially sensitive content that have the potential to cause harm by being witnessed, such as images of graphic violence, synthetic data can be deployed for two specific objectives. Its first role involves confining the handling of sensitive content to authorised individuals only, thereby curtailing the wider spread of this potentially harmful data. The second function is in the development of accurate models that simulate vital characteristics of the content intended to train automated online safety detection systems. Using synthetic data to replace such sensitive contetn in AI model training means that only trained specialists from law enforcement agencies can remain in charge of managing the original sensitive materials, hence limiting the amount of people who have access to them. The creation of such training datasets shows promise as it might limit the number of professionals that have access to the original content (Laranjeira da Silva et al. 2022).

## Provenance and Explainability

The provenance of synthetic data is a key issue. In the past, the metadata associated with synthetic data has been criticised for becoming out of date or incomplete (Cheah & Plale, 2012). This problem becomes especially acute if synthetic datasets are used to create models that themselves generate more synthetic data, leading to a potentially long feedforward chain which becomes more and more distant from the original dataset. Thorough processes of tracing provenance in data modelling should be applied with special care in the generation of synthetic data (Mu et al., 2022). In this regard, the Information Commissioner's Office recommends that AI development teams adhere to a transparent and diligent documentation process, recording "when and how" synthetic data was created and detailing the properties of synthetic data to both explain and justify appropriateness of decisions taken around model training (ICO, 2022). Another notable solution is the open sourcing of synthetic data algorithms. To this end, Massachusetts Institute of Technology's (MIT) Data to AI Lab has created the Synthetic Data Vault Project[10], an open source Python library of algorithms to generate synthetic data, stated to have been used over one million times (Ghoshal, 2022).

However, challenges to tracing provenance remain, especially when considering the use of foundational models (such as large language models), that have been pre-trained on large, generic datasets, as the basis for synthetic data creation. With the use of data such as text and images, there is an increasing trend towards user-involvement and user-prompted data generation using generative models like ChatGPT or DALL·E 2. Such models need to take into account user involvement in ensuring explainability, to promote transparency, trust, bias mitigation, and privacy awareness (Mohseni et al., 2021). There is also the question of the explainability of synthetic datasets once generated. Of course, one potential use case of synthetic data is to improve model explainability: by fabricating data and then using it to generate different types of predictions from a model, it is possible to explore how model outcomes relate to input data in a privacy preserving way (Papenbrock & Ebert, 2022).

# Sharing synthetic data

## Privacy concerns of sharing synthetic datasets

A core use case and potential benefit of synthetic data is their potential to enable data sharing in a privacy protecting way. Such sharing has immense potential practical benefits in the online safety space. For example, if social media platforms can safely share examples of self-harm, or text relating to bullying, then researchers and third parties can potentially help develop the models and technology required to automatically detect such harms, enabling more innovation in this area. Such sharing could also allow more industry wide benchmarks to be developed, mitigating concerns that the majority of models are trained and benchmarked on just a few publicly available datasets that may not accurately reflect the wider social context in which the models are then expected to operate (Kiela et al., 2021).

---

[10] **https://sdv.dev/**

> It must be noted that, synthetic datasets are not necessarily privacy preserving. Generative models, even when designed to create simulated data, may reproduce fragments of the real data they were trained on in the process of generation. This type of 'leakage', as it is sometimes known, is a generic problem across machine learning that is already well studied in the case of predictive models (Veale et al., 2018; Yeom et al., 2018), and is now also demonstrated in the case of more recent generative models by both academic studies (Carlini et al., 2021; Chen et al., 2020) and investigative reporting (Heikkilä, 2022).

The field has generated some suggestions in terms of improving privacy in synthetic datasets. One suggestion is to create synthetic datasets that are as narrowly targeted to the potential end use case as possible, rather than creating generic datasets that can be used in multiple ways. Such narrow datasets with fewer parameters within them should be more privacy preserving (Noble, 2022). Another possibility is to attempt to embed differential privacy in the creation of synthetic data sets. Differential privacy is a technique used to analyse and generate data in a way that protects individual privacy while still allowing useful information to be represented in the dataset. It involves preventing any individual data points of a single record (e.g. of details relating to an individual) from being distinguishable or identifiable, while still allowing general trends to be observed across the dataset (e.g. distribution across a population or group of people). Differential privacy therefore reduces any potential loss of utility by retaining overall statistical representativeness of the dataset, as well as ensuring higher levels of privacy, reducing leakages and the chances of re-identification (Abadi et al., 2016; Bellovin et al., 2019). These techniques are not necessarily a panacea, and some researchers doubt whether the degree of privacy can ever be high enough within a synthetic dataset, especially one where fidelity is not sacrificed (Stadler et al., 2022). However, more importantly from our perspective, they have been developed largely with tabular data in mind: it is not necessarily straightforward to apply them directly to other modalities such as text and images.

A further problem in this respect is that measurement of the extent of data leakage in a synthetically generated dataset is not straightforward. In the case of tabular data, many studies approach this problem by measuring the similarity of generated records against existing ones (Abowd & Vilhuber, 2008). However, this approach has recently been criticised as underestimating the potential for leakage (Stadler et al., 2022). New frameworks for measuring privacy in tabular data are starting to emerge (Houssiau et al., 2022). Meanwhile, the concepts of 'similarity' in the case of text and image data present even more measurement difficulties. For example, some applications have suggested examining the overlap between synthetic documents and real text on the basis of sequences of words (Libbi et al., 2021), whilst others have taken a differential privacy approach looking at whether model outputs are changed by removing individual instances of the training data (Melamud & Shivade, 2019). Other measures of similarity include measures of "embedding distance" (a numerical representation of text which should hopefully account for text with similar meaning being expressed differently - see e.g. (Zhao et al., 2019)) and planting "canaries" in the training data to see if they can be reconstructed in the generation phase (Yue et al., 2023). However, there exists no settled consensus on the best way of assessing data leakage and it remains an area of active academic investigation, with many of the metrics above often corresponding poorly to actual human judgement (Sai et al., 2020).

## Will sharing synthetic data (or models) on harms be used to generate harms?

A further concern worth mentioning is the use of synthetic data generation techniques by malicious actors. There are already concrete examples of generative models being used to create online harms once released. For example, one researcher trained a generative model on data from the online discussion site 4chan, which is well known for being very lightly moderated and as such home to a considerable amount of racist abuse and hate speech (Zelenkauskaite et al., 2021). The model itself was then allowed to post on 4chan, creating around 15,000 posts in the day it was active that were 'just as vile as the posts it was trained on' (Gault, 2022). More recently, research continues on the extent to which the next generation of large language models can be used to create harmful content: to take just one example, researchers at CheckPoint claimed to have discovered instances of malware being improved through the use of the recently released ChatGPT (CheckPoint, 2023). These concerns are becoming particularly important as synthetic data generation starts to become indiscernible from naturally generated data of human provenance.

> In a study on synthetically generated disinformation, (Spitale et al., 2023) found that humans cannot discern the difference between tweets produced by GPT-3 and those produced by other humans, and indeed that GPT-3 generated tweets can produce more compelling and truer-sounding disinformation than organic tweets produced by humans.

Within the online harms space, there has always been a concern about the balance between releasing data for research purposes with the intent of combating harms, and the potential risks posed by the actual data being released (see, for example, the controversy over the 'Jihadology' website, a source intended as a research repository of Jihadi related terrorist material that was then accused of facilitating its dissemination (Bond, 2018). However, these examples make it clear that there is a need to consider how to control the release of both synthetic datasets (that could be used to train generative models of online harms) and the release of generative models that can be used to create these synthetic datasets. The best way of doing this is unclear, and at the moment control around such models is highly dependent on the organisation releasing them. For example, ChatGPT is available as an API, but the model itself has not been released: and the model that can be queried through the API has in-built safeguards that aim to stop its potential use for the creation of harmful content (though there are by now numerous examples of the model being 'jailbroken' and escaping such safeguards). By contrast, Meta's first LLaMA model was leaked via 4chan (Vincent, 2023), meaning that there is no real meaningful control over its potential onward training and re-use[11] (and indeed at the time of publishing there were several other examples of 'open source' language models either already available or set to be released).

# Using synthetic datasets

## Defining Data Quality

When using synthetic datasets it can be challenging to assess quality, and ensure its representativeness of the underlying dataset. For example, low probability events in a naturally occurring dataset could be missed in a synthetically generated counterpart, especially when privacy preserving aspects of the synthetic generation algorithm may have a tendency to pull outliers more towards mean values (UK Statistics Authority, 2022). Preserving the privacy of very unusual cases is also potentially more problematic. A partial solution to this is to try and promote the generation of outliers that have not been seen in an original dataset, something which has proven to increase performance in computer vision (Tremblay et al., 2018). In response to the demand for an increased understanding and compatibility in synthetic data quality, the MIT's Data to AI Lab has also started DataCebo[12], a second open-source Python library focusing on evaluating model-agnostic tabular synthetic data for metrics of statistics, efficiency and privacy of data.

In contexts where personal data is involved, it is beneficial to review the aims of how to define the 'quality' of a synthetic dataset. In a modelling context, inappropriate quality measures could result in synthetic data not performing appropriately when deployed in real-world contexts. For instance, imagine a context where synthetic image data are created as a training dataset for models to improve their identification of people from images, for example to do an initial screening on a website accessible to children to see if the material might be pornographic. A challenge raised here is that where the dataset may be of high quality defined according to the clarity or usability of the images, but the success of the model is dependent on the number of visual features of the human body and clothing captured in the dataset, as well as the and demographic diversity in the images (Delussu et al., 2022).

There are a number of metrics that might prove helpful in evaluating the quality of a synthetic dataset. Evaluation metrics should be decided on a case-by-case basis, and based on what constitutes success for each individual application or use of synthetic data.

---

[11] Since the leak of LLaMA, Meta and Microsoft have jointly released LLaMA2. See: **https://ai.meta.com/llama/**
[12] **https://datacebo.com/**

- A **distribution comparison**, such as a Z-test or a two-sample Kolmogorov–Smirnov test is a helpful measure to compare each variable in the synthetic dataset with the equivalent variable in the real dataset it sets out to replicate (James et al., 2021; Tran-The, 2022), though this is more directly applicable to the case of tabular data.

- **Hellinger distance** is a measure that can be used to compare the distance between multivariate distributions in the real and synthetic dataset. This is a probabilistic measure between 0 and 1, where 0 indicates no difference between distributions (El Emam et al., 2020; James et al., 2021). Again, this applies most directly to tabular data.

- **Prediction accuracy** can be used to assess the utility of a synthetic dataset, whereby one compares the synthetic data's ability to replicate a prediction performed on the real dataset. This metric is particularly helpful if one does not have access to the original dataset, but instead predictions or analysis derived from it (James et al., 2021; Tran-The, 2022).

- Another way to assess performance without having the original dataset is by computing the **area under the receiver operating characteristic (AUC-ROC)**. AUC-ROC plots the specificity and sensitivity by comparing the relation between true positive and false positive rate (El Emam et al., 2020).

- One can model for **distinguishability** by shuffling the synthetic and real data and assess how hard it is for a machine learning model to distinguish between the real and synthetic instances. Perfect synthesis would mean that it is impossible for the model to predict the datapoint's true label (El Emam et al., 2020).

- A **bivariate correlation**, such as a Pearson's correlation, indicates whether there is a statistically significant linear relationship between two continuous variables. A stronger relationship indicates higher similarities between the synthetic and real dataset (James et al., 2021).

- Synthetic text can be assessed in respect to its **novelty**, measured by the lexical difference, such as Jaccard similarity, to the real dataset it is modelled off of. Similarity, diversity can be measured by the **repetition rate** of terms (ngrams) within a corpus (Bonaldi et al., 2022).

- The efficiency metric **HTER** can also help evaluate the quality of synthetic text by measuring the post-editing effort of an annotator. Generally, a value higher than 0.4 is used as a cut-off for poor quality text, where it is estimated that rewriting it from scratch would require as much effort as correcting it (Turchi et al., 2013).

Broadly speaking, these metrics can be divided into two categories. The first is resemblance metrics, that seek to understand whether the synthetic data preserves key characteristics and distributions within the data. The second, utility metrics, assess the data on the basis of whether it is possible to reach similar results using the synthetic data as one does using its real counterpart. However, some of the metrics assume tabular data structures. Additional metrics, as well as qualitative evaluations, will be needed as the use of synthetic image, video and audio data, and multimodal combinations of this content, increases. Note that there are separate metrics to assess the quality of synthetic data generating models (see (Dankar & Ibrahim, 2021; El Emam et al., 2022; Xu et al., 2018).

## Addressing bias and data gaps with synthetic datasets

In addition to increasing the availability of data (and potentially making it easier to share) one of the key potential benefits of the synthetic data approach is in the area of mitigating bias present in existing datasets (Verma et al., 2019). The existence of bias in data, and the way such bias can result in discriminatory models when trained on such data, is of course a well known problem within the field of AI and machine learning (Mehrabi et al., 2021). Adding synthetic data points that represent specific harms to under-represented groups such as marginalised communities could be particularly beneficial to improve model performance as these data points tend to be scarce. Consequently, a model performing worse due to having insufficient data might lead to inconsistent harm detection related to certain demographics or features distinct to a marginalised or under-represented community.

While mitigating bias in training data is not a guarantee that downstream models will produce unbiased outcomes, it is a potentially important step. Such bias mitigation could, for example, involve increasing the representation of a previously underrepresented group within the new synthetic dataset, or the generation of previously unseen examples (Jaipuria et al., 2020). To give a concrete use case from the domain of online safety, a toxicity detection model was trained using synthetic tweets that had been "translated" into an alternative phrasing with African American English dialectal marker (Zhou et al., 2021). Such novel data potentially makes it possible to evaluate the effectiveness of models made to capture online harms in contexts for which there is not original data available.

> It is worth emphasising that, while the generation of synthetic data to mitigate bias in datasets is a promising research avenue, generative models for creating synthetic data are likely biased themselves, experiencing so called 'bias creep' from the data they are trained on (Feng et al., 2023). Such biases have been observed, for example, in the case of image generation models reproducing negative stereotypes (Bianchi et al., 2022), whilst it has been argued that ChatGPT has a 'left-libertarian' bias (Hartmann et al., 2023).

Hence, careful attention to potential bias in generated synthetic datasets is important. When mitigating for bias, regardless of field or domain, it is also important to have a good understanding of the topic in question. Some types of data might have a natural skew toward a specific demographic, and having domain knowledge of the expected distribution is important to be able to assess where there are true gaps in the data, versus data simply reflecting a natural skew. When mitigating for bias, one should be mindful of the intended use-case, as increasing the representation within the data could, for example, enable the new models to detect outliers that otherwise would not have been captured in the natural distribution whilst the same data, instead used for prediction, would render misleading results.

# Conclusions

This report has explored the opportunities presented by the use of synthetic data in the context of developing technologies to help improve online safety, and outlined the main issues that may arise from doing so. Our research suggests that synthetic data present many opportunities to address some of the prominent limitations related to the responsible collection, storage, and use of large real world datasets. However, similar to models trained on authentic data, models trained on synthetic data also risk degradation over time, meaning that they may not capture new forms of harm or their manifestation in new contexts. Using synthetic data allows researchers and data scientists to mitigate bias in data and simulate relevant hypothetical scenarios, and may streamline responsible data sharing by creating more effective privacy-preserving methods. Synthetic data can also help with increasing the volume of data available (e.g., augmenting dataset), and also with enlarging the diversity of examples, in turn helping mitigate bias. This is especially important in online safety applications, where data points of specific harms to particular groups of people may be insufficient in quantity, leading to potential disparities in harm detection across different groups.

Given its potential as a solution to help many of the most significant limitations of existing 'genuine' datasets, as well as overcoming many technical barriers to entry for research and development - there has been a growing interest in using synthetic data across a wide range of applications. However, it also poses novel, often less explored, ethical challenges which must be considered when creating, sharing, and using synthetic data. One of the biggest promises of synthetic data is its potential as a privacy preserving alternative to real-world data. But, synthetic data is not necessarily privacy preserving, and more importantly, there exists a relevant trade-off between privacy and fidelity (and thus utility) in synthetic datasets. In particular, as synthetic data veers farther from the original dataset, increasing its privacy, it also tends to lose its utility as a means to study dynamics pertaining to the original dataset. Additionally, the privacy level of synthetic data may not be obvious because measuring 'data leakage' in synthetic datasets is not a straightforward task. This is particularly problematic when measuring 'similarity' in non-tabular data, as two data points may differ substantially between them yet still convey the same message across. As such, assessing the levels of privacy of synthetic datasets used for online harms is a complex issue which must be explored before their widespread adoption.

The sharing of synthetic datasets in the context of online harms research and development also remains sensitive, because if freely and openly shared, synthetic datasets created to study online harms might then be used by malicious actors to create further harms. The generative component of synthetic dataset creation compounds this risk; for any one example of an online harm, several potentially new examples of harms may be created (and disseminated). Furthermore, even if bias mitigation can be improved using synthetic data, the problem will not necessarily be solved, especially because the models used to create synthetic data might be biassed themselves. Likewise, like the use of synthetic data cannot fully address the sociohistorical biases and discriminatory patterns which are often baked into training data and AI model inferences, because such technology-based solutions do not confront the underlying social, historical, and cultural causes of bias, inequity, and discrimination that manifest in data-intensive algorithmic systems.

Taken together, the potential risks and ethical hazards that accompany the application of generative AI models to create synthetic data, indicate that the use of such models, and the data they produce, should not be seen as a silver bullet that will, by itself, enable online safety through technological intervention. Careful evaluation of these generative models and their synthetic data output, as well as their applied use cases is crucial, but the potential opportunities they present to accelerate technological developments may well provide benefits that outweigh their risks if used responsibly.

turing.ac.uk
@turinginst

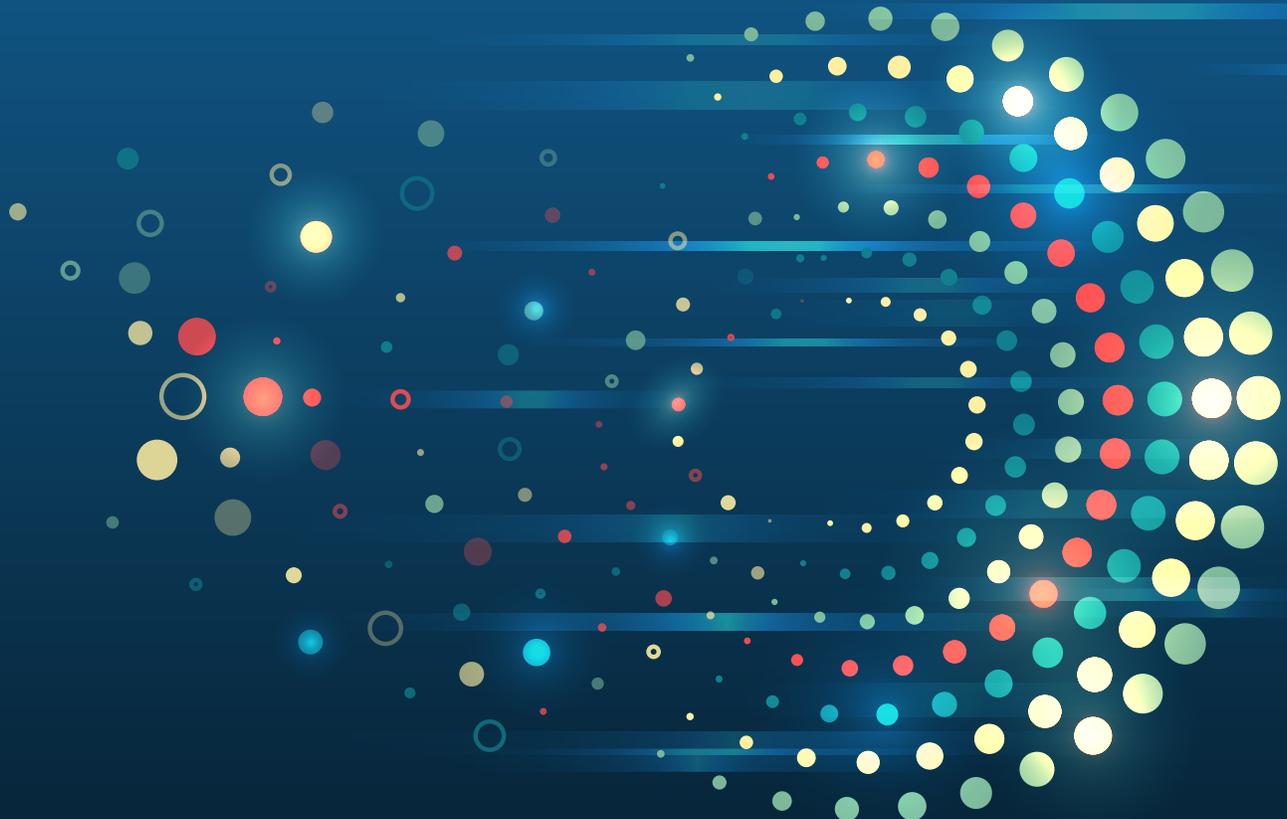